\documentclass[preprint,showpacs,preprintnumbers,amsmath,amssymb,nofootinbib,showkeys,aps]{revtex4}
\usepackage{graphicx}% Include figure files
\usepackage{dcolumn}% Align table columns on decimal point
\usepackage{bm}% bold math
\usepackage{latexsym}
\usepackage{epsfig}
\usepackage{hyperref}
\usepackage[english]{babel}

\newcommand{\mnras}{{\it Monthly Notices of Royal Astronomical Society}}
\newcommand{\jcap}{Journal of Cosmology and Astroparticle Physics}
\newcommand{\be}{\begin{equation}}
\newcommand{\ee}{\end{equation}}
\newcommand{\beqq}{\setlength\arraycolsep{2pt}\begin{eqnarray}}
\newcommand{\eeqq}{\vspace{0cm} \end{eqnarray}}
\newcommand{\bea}{\begin{eqnarray}}
\newcommand{\eea}{\end{eqnarray}}

\newcommand{\lPl}{\ell_{Pl}}

\begin{document}

\title{Thermodynamic constraints on matter creation models}
\author{R. Valentim}\email{valentim.rodolfo@unifesp.br}
\affiliation{Departamento de F\'{\i}sica, Instituto de Ci\^encias Ambientais, Qu\'{\i}micas e Farmac\^euticas - ICAQF, Universidade Federal de S\~ao Paulo (UNIFESP)
Unidade Jos\'e Alencar, Rua S\~ao Nicolau No. 210,
09913-030 -- Diadema, SP, Brazil}

\author{J. F. Jesus} \email{jf.jesus@unesp.br}
\affiliation{Universidade Estadual Paulista (UNESP), C\^{a}mpus Experimental de Itapeva \\
Rua Geraldo Alckmin 519, 18409-010, Vila N. Sra. de F\'atima, Itapeva, SP, Brazil}
\affiliation{Universidade Estadual Paulista (UNESP), Faculdade de Engenharia de Guaratinguet\'a \\
Departamento de F\'{i}sica e Qu\'{i}mica, Av. Dr. Ariberto Pereira da Cunha 333, 12516-410 - Guaratinguet\'a, SP, Brazil}
%\keywords{dark matter theory, particle physics - cosmology connection }

\begin{abstract}
Entropy is a fundamental concept from Thermodynamics and it can be used to study models on context of Creation Cold Dark Matter (CCDM). From conditions on the first ($\dot{S}\geq0$)\footnote{Throughout the present work we will use dots to indicate time derivatives and dashes to indicate derivatives with respect to scale factor.} and second order ($\ddot{S}<0$) time derivatives of total entropy in the initial expansion of Sitter through the radiation and matter eras until the end of Sitter expansion, it is possible to estimate the intervals of parameters. The total entropy ($S_{t}$) is calculated as sum of the entropy at all eras ($S_{\gamma}$ and $S_{m}$) plus the entropy of the event horizon ($S_h$). This term derives from the Holographic Principle where it suggests that all information is contained on the observable horizon. The main feature of this method for these models are that thermodynamic equilibrium is reached in a final de Sitter era. Total entropy of the universe is calculated with three terms: apparent horizon ($S_{h}$), entropy of matter ($S_{m}$) and entropy of radiation ($S_{\gamma}$). This analysis allows to estimate intervals of parameters of CCDM models.
\end{abstract}

%
%\flushbottom
%\pacs{}
\maketitle

%====================================================================================================================
%====================================================================================================================

\section{\label{introduction} Introduction}

%Physical systems tend spontaneously to reach thermodynamic equilibrium when energy is exchanged between a physical system and its neighbourhoods. This is the empirical basis of the second law of Thermodynamics. The Laws of Thermodynamics state that the entropy ($S$) for closed systems remain constant or increase with time ($\dot{S}\geq0$). The second order time derivative ($\ddot{S}<0$), at least roughly, leads to thermodynamic equilibrium \cite{Callen1985,MimosoPavon2013}. One way of imposing the condition on second order derivatives in cosmic expansion is through the Holographic Principle proposed by \cite{t'Hooft1993} and \cite{Susskind1995}, with direct application in Cosmology \cite{FischlerSusskind1998,BakRey2000}. This principle assumes that all information is on the the Universe horizon.

When physical systems are isolated they tend spontaneously to reach thermodynamic equilibrium. This idea is at the empirical basis of the Second Law of Thermodynamics: that the entropy ($S$) for closed systems remain constant or increase with time ($\dot{S}\geq0$). The second order entropy derivative with respect to the relevant variable must obey $\ddot{S}<0$, at least roughly,  when the Universe keeps to expand on the infinite future. It leads to thermodynamic equilibrium \cite{Callen1985,MimosoPavon2013}. One way of assuming the condition on second order derivatives in cosmic expansion is through the Holographic Principle proposed by \cite{t'Hooft1993,Susskind1995} that was directly applied in Cosmology \cite{FischlerSusskind1998,BakRey2000}. This principle assumes that all information is on the Universe horizon surface.

The matter creation in the context of cosmology has been studied by different authors. Ref. \cite{Parker1968} investigated particle creation mechanisms using covariant quantized free field equations of elementary particles in the expansion of the Universe. In this work, the author analysed the creation of particles with spin--0 pions, spin--1/2 and particles with zero mass and non-zero rotation.

Prigogine {\it et al.} \cite{PrigogineEtAl1988} argue that Einstein's equations for General Relativity are adiabatic and reversible, so they do not allow the production of entropy in a cosmological scenario. The authors proposed a way to solve this problem based on the idea of irreversibility of thermodynamic systems. Authors showed that the Thermodynamics of irreversible systems leads naturally to a reinterpretation of Einstein's equations, which allows the creation of matter from the gravitational field and consequently the production of entropy. The cosmological history proposed by \cite{PrigogineEtAl1988} has three stages: first, from an initial vacuum fluctuation in de Sitter's space; second, that de Sitter space exists during a time of decay of its constituents; third, a phase transition transforms this de Sitter space into a universe with a Friedman-Robertson-Walker metric that evolves adiabatically on the cosmological scale. An important aspect to be emphasized is that the approach of the matter creation does not consider Dark Matter and Dark Energy scenarios. A natural consequence of this approach is the rate of change in the number of particles, $\Gamma\simeq0 $, which spells out the Second Law of Thermodynamics. There are still unanswered questions about the creation mechanism ($ \Gamma $) from the gravitational field, the physical nature of the particles and how $ \Gamma $ influences the expansion of the universe \cite{PanEtAl2016}. Some authors suggest that the type of particles created in this process are limited by local links related to gravity \cite{EllisEtAl1989, HagiwaraEtAl2002, PeeblesRatra2003}. These authors showed that radiation does not contribute significantly to the late accelerated expansion of the universe in the dominant dark matter phase. Some later work suggests that the particles produced by the gravitational field are Cold Dark Matter (CDM) particles and that for the rates of matter creation ($ \Gamma $) may constrain $\Lambda$CDM \cite{SteigmanEtAl2009, LimaEtAl2010, FabrisEtAl2014, LimaEtAl2014, ChakrabortyEtAl2015}.

In a recent work \cite{NunesPavon2015}, the authors present the possibility of a quantum vacuum equation of state associated with the creation of particles by the gravitational field that acts in a vacuum. They analyzed three different matter creation rates $ \Gamma $ and estimated the parameters from SNe Ia, Gamma Ray Bursts (GRBs), Baryon Acoustic Oscillations (BAO) and Hubble parameter data. The authors show that matter creation models can explain the phantom behaviour of our Universe without the need to insert phantom fields \cite{Cadwell2002}. The work proposed by \cite{PanEtAl2016} analyses how the process of matter creation happens with the universe expanding. In the context analysed by the authors, the gravitational field induces a process of adiabatic matter creation. In this work, \cite{PanEtAl2016} present a generalized model for $ \Gamma $ with three free parameters $ \Gamma = \Gamma_0 + l H
^2 + nH + m/H $ \cite{PanChakraborty2015, ChakrabortyEtAl2014}. This model encompasses the transition from the inflationary phase to the radiation phase for adiabatic particle production. In another recent work \cite{PanEtAl2019}, it is proposed a two-fluid model where one fluid ($ \rho_1 $) is produced adiabatically and there is another fluid that does not interact with fluid 1 and satisfies the energy conservation equation. One important aspect of the work is the study of the singularities of $\Gamma$ from the analysis of a series expansion. With this they plot the profiles of $ q $ according to the scale parameter ($ a $) for each $ \Gamma $ in relation to the terms of the expansion of $ \Gamma \equiv \Gamma(H)$. Although interesting the idea of a series expansion, we shall restrict ourselves here to the analysis of a simpler and yet broad class of matter creation models.

We will explore in this work the calculations of the total entropy ($S$) from the holographic principle for five models of matter creation. These models were studied by \cite{bic-ccdm} and it assumes that the creation rate $\Gamma$ is a function of the Hubble parameter $H$. Each dark matter creation rate leads to a different cosmic evolution \cite{GraefEtAl2014,Freaza2002,Lima2008,JO16,Lima2010,LimaBasilakos2012,LimaBasilakos2013}\footnote{See also \cite{JesusPereira14,LimaBaranov14} for more fundamental formulations of matter creation models.}. A common feature of these models is that the Universe starts  in an inflationary, de Sitter phase, then it passes through the ages of radiation and matter, where it finally enters the final de Sitter stage. Total entropy ($S$) at each phase is equal to the sum of each entropy contribution for these different ages. $S$ is the direct sum of the contribution of entropy to radiation, matter and the apparent horizon of the Holographic Principle \cite{t'Hooft1993,Susskind1995}:
\be
S = S_{\gamma} + S_{m} + S_{h};
\label{Stot}
\ee
where $S_h=\frac{k_B\mathcal{A}}{4\ell_{Pl}^2}$, is the entropy of the apparent horizon, $S_m$ is entropy of pressureless matter and $S_\gamma$ is entropy of radiation. $\mathcal{A}$ and $\ell_{Pl}$ denote the area of the horizon and Planck's length, respectively. In an ever expanding Universe, the conditions $\dot{S}(t)>0$, $\ddot{S}(t\rightarrow\infty)<0$ are equivalent to the conditions $S'(a)>0$, $S''(a\rightarrow\infty)<0$. Restricting our analysis to this class of models, we shall consider the entropy as a function of the scale factor from now on.

In this work, entropy evolution will be considered, initially based on the model proposed by \cite{LimaBasilakos2012,LimaBasilakos2013} and the models analyzed by \cite{bic-ccdm}. We can use the conditions on the derivatives of the total entropy to estimate the intervals of validity of free parameters for each model \cite{MimosoPavon2013}. We shall assume a FRW metric, in agreement with the Cosmological Principle, and a spatially flat universe, as predicted from most inflationary models. However, recently, Ref. \cite{DiValentino2019} have shown from Planck Legacy 2018 dataset analysis that the curvature parameter $\Omega_k\equiv-k/(a_0^2H_0^2)$ can have a non-zero value, namely, $-0.095<\Omega_k<-0.007$ at 99\% c.l.
As this ``new'' value of $\Omega_k$ is a controversial theme, and the deviation from spatial flatness seems to be small, we prefer to use the standard value in this work, as predicted by most inflationary models. So, we are restricting our analysis to spatially flat models ($k=0$).

\section{Creation of Cold Dark Matter Models (CCDM)}

Models of CCDM used in this work it were statistically analyzed by \cite{bic-ccdm} and have a natural dependence of $H$ ($\Gamma\equiv\Gamma(H)$), where $\Gamma$ as function of Hubble parameter represents a relation between the matter creation and expansion rates. All the CCDM models used here have also free parameters.  The models studied here were analyzed by \cite{bic-ccdm} using three statistical criteria: Bayesian Information Criterion (BIC), Akaike Information Criterion (AIC) and Bayesian Evidence (BE) using the SNe Ia dataset. Most of these models can be described by a function $\Delta=\beta E+\alpha E^{-n}$, where $\Delta\equiv\frac{\Gamma}{3H}$ and $E\equiv\frac{H}{H_0}$. So, it corresponds to a creation rate $\Gamma=3\beta H+3\alpha H_{0}\left(\frac{H_{0}}{H}\right)^n$.

Another model analyzed in \cite{bic-ccdm} is LJO \cite{ljo10} with $\Gamma=3\alpha\frac{\rho_{c0}}{\rho_{dm}}H$. The LJO model has the same dynamics as $\Lambda\mbox{CDM}$ concordance model. In LJO, the cosmological constant is exactly mimicked by particle creation. Due to this mimicking, we choose not to analyze this model here, as $\Lambda$CDM has already been thoroughly analyzed on \cite{RadPav12}. In all models analyzed in this work we have neglected the contribution of baryons. The baryonic contribution is small, $\sim5\%$ of Universe content and our results can be more dependent on the assumptions made here in order to estimate entropy rather than baryonic influence. Another important assumption is that Universe is spatially flat as indicated from CMB and preferred by inflation, i.e. $\Omega_k\equiv 0$ in our analysis. The models studies here are described on Table \ref{models}.

\begin{table}[ht]
\renewcommand{\tabcolsep}{1.5pc} % enlarge column spacing
\renewcommand{\arraystretch}{1.2} % aumenta line spacing
	\centering
		\begin{tabular}{cccccc}
\hline\hline
Model &  Creation rate & Reference & Parameters\\
\hline

$M_1$	& $\Gamma=\frac{3\alpha H_0^2}{H}$ 									& \cite{JO16} (JO)																& $\beta=0$, $n=1$\\
%$M_1$ & $\Gamma=3\alpha\frac{\rho_{c0}}{\rho_{dm}}H$ 			& \cite{ljo10} (LJO)% ($\sim\Lambda$CDM)
 %	& $\alpha\in[0,1]$\\
$M_2$	& $\Gamma=3\alpha H_0$ 															&	\cite{GraefEtAl14}									& $\beta=0$, $n=0$\\
$M_3$	& $\Gamma=3\beta H$ 																&		--							& $\alpha=0$\\
$M_4$	& $\Gamma=3\alpha H_0\left(\frac{H_0}{H}\right)^n$ 	&			--					& $\beta=0$\\
$M_5$	& $\Gamma=3\alpha\frac{H_0^2}{H}+3\beta H$ 					&	\cite{GraefEtAl14}									& $n=1$\\
\hline
\hline
		\end{tabular}\\
\caption{Models and parameters.}
\label{models}
\end{table}

\section{Methodology}

The methodology adopted here consists on analyzing total entropy of the Universe in the context of matter creation models. This analysis allows to estimate the validity interval for free parameters for each model. This idea is based on \cite{MimosoPavon2013}, where authors analyzed first and second order derivatives. It assumes the Second Law of Thermodynamics jointly with the idea that thermodynamic equilibrium must be achieved at some future time. An important aspect of this method is that it takes into account the horizon entropy that came from Holographic Principle \cite{t'Hooft1993,Susskind1995, FischlerSusskind1998} where all the information about Universe is on horizon. The total entropy is given by equation (\ref{Stot}) and it is defined as sum of radiation, matter and apparent horizon. Restricting our analysis to CCDM models \cite{Calvao1992,Lima2012,LimaBasilakos2013,bic-ccdm,GraefEtAl14}, entropy was considered as a function of the scale factor. In CCDM, expansion acceleration can be achieved through an effective creation pressure:
\be
p_c=-\frac{(\rho+p)\Gamma}{3H}=-\frac{\rho\Gamma}{3H};
\ee
where $p_c$ is creation pressure, $\rho$ is dark matter (DM) density (pressure $p$ vanishes for DM), $\Gamma$ is creation rate and $H$ is Hubble parameter. Relation between Hubble parameter and $\rho$ is the Friedmann equation:
\be
H^2=\frac{8\pi G}{3}\rho.
\label{fried}
\ee
for spatially flat Universe ($k=0$). The equation of continuity for dark matter now reads:
\be
\dot{\rho}+3H\rho=\Gamma\rho,
\ee
That is $\Gamma\rho$ is a source ($\Gamma>0$) or sink ($\Gamma<0$) for dark matter. % Note that we use the dot notation to express the derivative with respect to time ($^\dot\equiv d/dt$).
The Hubble parameter corresponds to the expansion rate, that is, $H=\dot{a}/a$, so, writing it as a function of scale factor, we have:
\be
%\rho'(a)+\frac{3}{a}\rho(a)=\frac{\Gamma}{aH}\rho(a)\Rightarrow
\rho'(a)=\frac{\rho(a)}{aH}(\Gamma-3H).
\ee
where we denoted the derivative with respect to $a$ with a prime. The relation between matter density $\rho$ and particle number density $n$ is $\rho=nm$, where $m$ is mass of DM particle, so, we have:
\be
%\rho=nm\Rightarrow 
n'=\frac{n}{aH}(\Gamma-3H).
\label{ncont}
\ee

The Friedmann equation and continuity equation fully describe the CCDM background dynamics. From these equations we can derive a relation between $H$ and $\Gamma$:
\be
\dot{H}+\frac{3}{2}H^2\left(1-\frac{\Gamma}{3H}\right)=0
\label{Hdot}
\ee
or, in terms of scale factor,
\be
H'=-\frac{3H}{2a}\left(1-\frac{\Gamma}{3H}\right)
\label{Hlinha}
\ee

This class of models suggests that matter creation ($\Gamma>0$) generates a negative pressure ($p_c<0$) which may explain the acceleration of the Universe.

\section{\label{model}Thermodynamics of Matter Creation Models}
In our analysis we are interested only on recent and future times, so we shall restrict ourselves to the matter dominated age, as radiation becomes negligible in the past. From equation (\ref{Stot}), shown earlier, we will analyze the derivatives of each of the terms for the total entropy: entropy of the apparent horizon, matter and radiation \cite{MimosoPavon2013}.

Entropy of apparent horizon is $S_{h}=k_{B}\mathcal{A}/(4l^2_{Pl})$, where $\mathcal{A}$ denotes the area of apparent horizon and $l_{Pl}$ is Planck's length. The area of the apparent horizon is given by $\mathcal{A}=4\pi \tilde{r}_\mathcal{A}^2$, where $\tilde{r}_\mathcal{A}=\frac{1}{\sqrt{H^2+ka^{-2}}}$. As explained above, we are restricting our analysis to spatially flat models ($k=0$). This assumption yields $\tilde{r}_\mathcal{A}=H^{-1}$ and $\mathcal{A}=4\pi H^{-2}$. In this case, the horizon entropy reads:
\be
S_h=\frac{k_B\pi}{\ell_{Pl}^2H^2};
\ee

That is, the entropy is function of Hubble parameter only. Thus, the first derivative of apparent horizon entropy with respect to scale factor is:
\be
S'_h=-\frac{2k_B\pi H'}{\ell_{Pl}^2H^3}.
\label{dShda}
\ee
The first-order derivative of the entropy results in an expression that is a function of $H$ and its first derivative. Eq. (\ref{Hlinha}) yields $H'=\frac{\Gamma-3H}{2a}$, thus we may write for $S_h'$:
\be
S'_h=\frac{k_B\pi}{\ell_{Pl}^2aH^3}(3H-\Gamma).
\label{dShda1}
\ee

For the $S_m$ entropy, we may consider that every single particle contributes to the entropy inside the horizon by a single bit, $k_B$ \cite{MimosoPavon2013}. In this case, we have:
\be
S_m=k_B\frac{4\pi}{3}\tilde{r}_\mathcal{A}^3n=k_B\frac{4\pi n}{3H^3},
\ee
where $n$ is the number density. By deriving this equation we find:
\be
S'_m=\frac{4\pi k_B}{3H^4}(n'H-3nH').
\label{dSmdaH}
\ee

This expression is first derivative of entropy as function of $H$, $H'$ and $n$. By using Eqs. (\ref{ncont}) and (\ref{Hlinha}), we may write:
\be
S'_m=\frac{2\pi k_{B} n}{3aH^4}(3H-\Gamma)
\label{dSmda}
\ee

That is, the derivative of entropy of matter as function of $H$, $n$ and $\Gamma$. Now combining Eqs. (\ref{dShda}) and (\ref{dSmda}), we have
\be
S'=\frac{k_B\pi}{aH^3}\left(\frac{1}{\lPl^2}+\frac{2n}{3H}\right)(3H-\Gamma)
\label{dSda}
\ee

So, a necessary and sufficient condition for having $S'\geq0$ is $\Gamma\leq3H$, that is, the particle creation rate must be less or equal to the volumetric expansion rate\footnote{Any volume $V$ in the Hubble flow scales with $a^3$, thus $\frac{\dot{V}}{V}=3H$.}. Let us define the dimensionless quantity $s_1$:
\be
s_1\equiv\frac{3H-\Gamma}{H_0}
\label{s1}
\ee

Thus, $S'\geq0$ corresponds to $s_1\geq0$. Now, let us impose the concavity condition $S''(a\rightarrow\infty)<0$, that is, impose that the Universe reaches thermodynamic equilibrium in the infinite future. By deriving (\ref{dShda}):
\be
S''_h=\frac{2\pi k_B}{\lPl^2H^4}(3H'^2-HH'')
\label{d2Shda2}
\ee

Using $\rho=nm$ and deriving the Friedmann equation (\ref{fried}), we have
\be
2HH'=\frac{8\pi Gn'm}{3}
\ee

Combining this with the Friedmann equation, we find the relation\footnote{It can also be found from Eqs. (\ref{ncont}) and (\ref{Hlinha}).}
\be
2\frac{H'}{H}=\frac{n'}{n}\Rightarrow2H'n=Hn'
\label{Hnrelation}
\ee

That is, the particle density relative variation (w.r.t. $a$) is double of Hubble parameter relative variation. We may use this to simplify  Eq. (\ref{dSmdaH}):
\be
S'_m=-\frac{4\pi k_BnH'}{3H^4}
\label{dSmda1}
\ee
Now it is easier to derive it to find $S''_m$:
\be
S''_m=\frac{4\pi k_Bn}{3H^5}(2H'^2-HH'')
\label{d2Smda2}
\ee
where we have derived (\ref{dSmda1}) and used the relation (\ref{Hnrelation}) again in order to omit $n$ derivatives. By summing (\ref{d2Shda2}) and (\ref{d2Smda2}), we find:
\be
S''=\frac{2\pi k_B}{\lPl^2H^4}(3H'^2-HH'')+\frac{4\pi k_Bn}{3H^5}(2H'^2-HH'')
\label{d2Sda2}
\ee

Let us define the dimensionless quantities:
\bea
s_{h2}\equiv\frac{3H'^2-HH''}{H_0^2}\label{sh2}\\
s_{m2}\equiv\frac{2H'^2-HH''}{H_0^2}
\label{sm2}
\eea

Thus, the conditions $S''_h<0$ and $S''_m<0$ correspond to $s_{h2}<0$ and $s_{m2}<0$, respectively. A sufficient (but not necessary) condition for having $S''<0$ is having both $s_{h2}<0$ and $s_{m2}<0$. Although it may be too restrictive a condition over the models, we consider it reasonable in order to achieve a result not much dependent on the choice of the contribution of each particle to the entropy ($S_m/N$).

Another interesting inference we can make from expressions (\ref{sh2}) and (\ref{sm2}) is that $s_{h2}=s_{m2}+\frac{H'^2}{H_0^2}$, so $s_{h2}\geq s_{m2}$ at all times, so every time that $s_{h2}<0$, we have $s_{m2}<0$. That is, $S''_h<0$ implies $S''_m<0$.

In the next section we will analyze a quite general model for the rate of creation of dark matter with three free parameters.

\section{Case Study: \texorpdfstring{$\Gamma=3\beta H+3\alpha H_0\left(\frac{H_0}{H}\right)^n$}{Gamma = 3 beta H + 3 alfa (H0/H)**n}}
We now analyze a quite general model of the matter creation rate which was derived by \cite{GraefEtAl14} with three free parameters: $ \alpha $, $ \beta $ and $ n $. All the models that we will deal with here are particular cases of this model whose dependence with $H$ is given by:
\be
\Gamma=3\beta H+3\alpha H_0\left(\frac{H_0}{H}\right)^n.
\label{gamageral}
\ee

This model for $\Gamma$ is a combination of two important dependencies: the first term $\propto H$ and the second term $\propto H^{-n}$. In this case, Eq. (\ref{Hlinha}) reads
\be
\frac{dE}{da}=\frac{3}{2a}\left[\alpha E^{-n}-(1-\beta)E\right]
\label{dEda}
\ee
where $E(a)\equiv\frac{H(a)}{H_0}$. As shown by \cite{bic-ccdm}, Eq. (\ref{dEda}) can be solved as
\be
E(a)=\frac{H(a)}{H_0}=\left[\frac{\alpha+(1-\alpha-\beta)a^{-\frac{3}{2}(n+1)(1-\beta)}}{1-\beta}\right]^\frac{1}{n+1},
\label{Hageral}
\ee
in case that $\beta\neq1$ and $n\neq-1$. Case $n=-1$ is equivalent to $\alpha=0$. If $\beta=1$, $E(a)$ can be obtained from (\ref{dEda}) as
\be
E=\left[1+\frac{3\alpha(n+1)}{2}\ln a\right]^{\frac{1}{n+1}}
%E=\left[1+3\alpha\ln a\right]^{1/2}
\label{Habeta1}
\ee

The eq. (\ref{Hageral}) shows $H(a)$ as a function of scale factor $a$, $H_0$, $\alpha$, $\beta$ and $n$. By writing $H(a)$ as an explicit function of the parameters, we can now impose the condition $S'\geq0$. From the Eq. (\ref{dSda}) and (\ref{s1}) it yields:
\be
s_1=3\left(\frac{H}{H_0}\right)^{-n}\left[(1-\alpha-\beta)a^{-\frac{3}{2}(n+1)(1-\beta)}\right]\geq0.
\label{Slingt0}
\ee

We must have $S'\geq0$ at all times, so we must have $1-\alpha-\beta\geq0$ by this analysis.

According to (\ref{d2Shda2}), $S''_h<0$ implies $3H'^2-HH''<0$, so
\be
\begin{split}
s_{h2}=\frac{3}{4}\left(\frac{H}{H_0}\right)^{-2n}\left(\frac{1-\alpha-\beta}{1-\beta}\right)\times\\
\times a^{-2-\frac{3}{2}(n+1)(1-\beta)}\left\{2(1-\alpha-\beta)(2-3\beta)a^{-\frac{3}{2}(n+1)(1-\beta)}+\alpha\left[3\beta-3n(1-\beta)-5\right]\right\}<0
\label{expSh}
\end{split}
\ee

Now, let us impose the condition $S''_m<0$. It implies, from (\ref{d2Smda2}) that $2H'^2-HH''<0$. We have
\be
\begin{split}
s_{m2}=\frac{3}{4}\left(\frac{H}{H_0}\right)^{-2n}\left(\frac{1-\alpha-\beta}{1-\beta}\right)\times\\
\times a^{-2-\frac{3}{2}(n+1)(1-\beta)}\left\{(1-\alpha-\beta)(1-3\beta)a^{-\frac{3}{2}(1+n)(1-\beta)}+\alpha\left[3\beta-3n(1-\beta)-5\right]
\right\}<0
\label{expSm}
\end{split}
\ee

%\be
%s_{h2}|_{a\rightarrow\infty}=
%\frac{3}{4}\left(\frac{H}{H_0}\right)^{-2n}\left(\frac{1-\alpha-\beta}{1-\beta}\right)\left\{\alpha\left[3\beta-3n(1-\beta)-5\right]a^{-2-\frac{3}{2}(n+1)(1-\beta)}\right\}|_{a\rightarrow\infty}<0
%\label{expShinf}
%\ee

%In order to impose the condition (\ref{expShinf}), we must have $1-\alpha-\beta>0$, $-2-\frac{3}{2}(n+1)(1-\beta)\geq0$ and $\frac{\alpha[3\beta-3n(1-\beta)-5]}{1-\beta}<0$.

We remind that we are interested in the sign of (\ref{expSh}) and (\ref{expSm}) only in the limit $a\rightarrow\infty$. However, this limit is strongly dependent in the parameter set $\{\alpha,\beta,n\}$, so, instead of putting limits for the general model, we shall put limits for each particular model. Let us do it in next subsections.

\subsection{\texorpdfstring{$M_1: \Gamma=\frac{3\alpha H_0^2}{H}$}{M1: Gamma = 3 alpha H0**2/H}}
In this case we have the fixed parameter values $\beta=0$ and $n=1$, so from (\ref{Slingt0}) we see that $S'\geq0$ reads 
\be
s_1=3(1-\alpha)\left(\frac{H_0}{H}\right)a^{-3}\geq0.
%\label{Slingt0}
\ee
which implies $\alpha\leq1$. From (\ref{expSh}), the condition $S''_h<0$ reads
\be
s_{h2}=3\left(\frac{H_0}{H}\right)^2\left(1-\alpha\right)a^{-5}\left[(1-\alpha)a^{-3}-2\alpha\right]<0
%\label{expSh}
\ee
Thus, for $a\rightarrow\infty$, it implies $0<\alpha<1$. From (\ref{expSm}), the condition $S''_m<0$ reads
\be
s_{m2}=\frac{3}{4}\left(\frac{H_0}{H}\right)^2\left(1-\alpha\right) a^{-5}\left[(1-\alpha)a^{-3}-8\alpha\right]<0
%\label{expSm}
\ee
which yields the same limit for $a\rightarrow\infty$, $0<\alpha<1$.

\begin{figure}[ht]
 \centering
 \includegraphics[width=.48\textwidth]{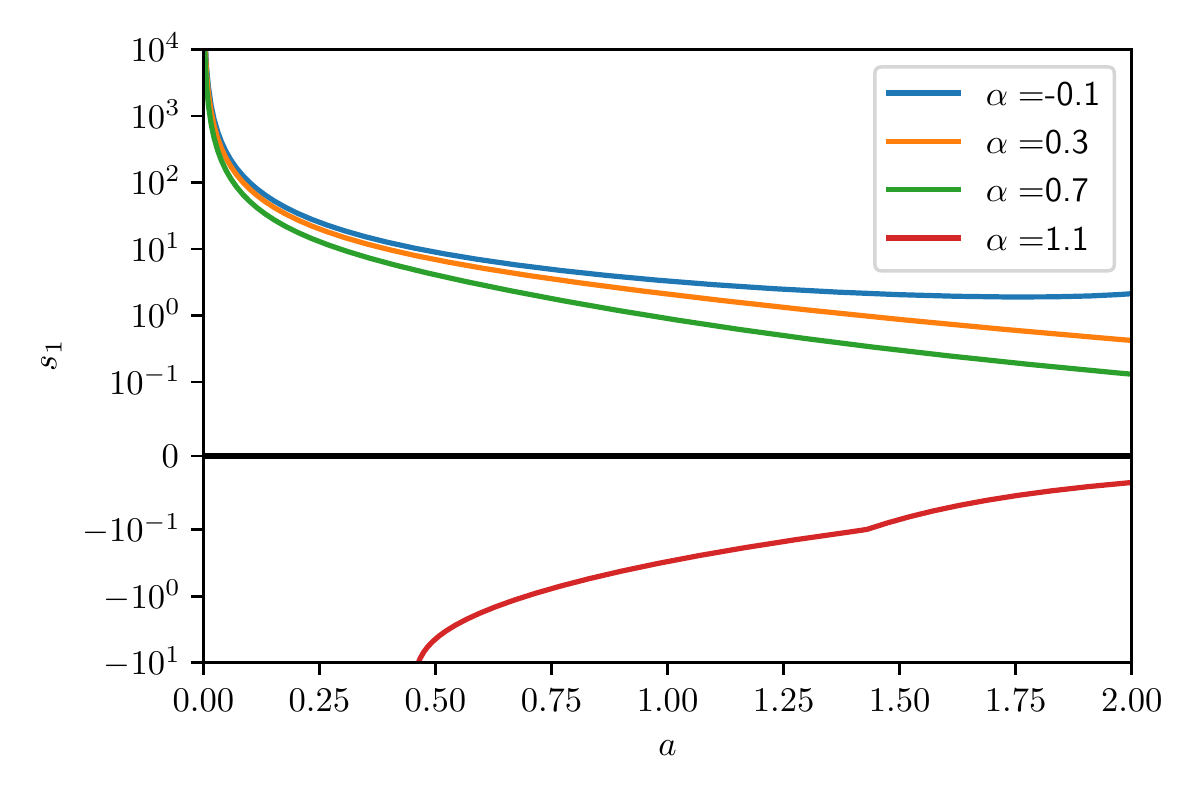}
 \includegraphics[width=.48\textwidth]{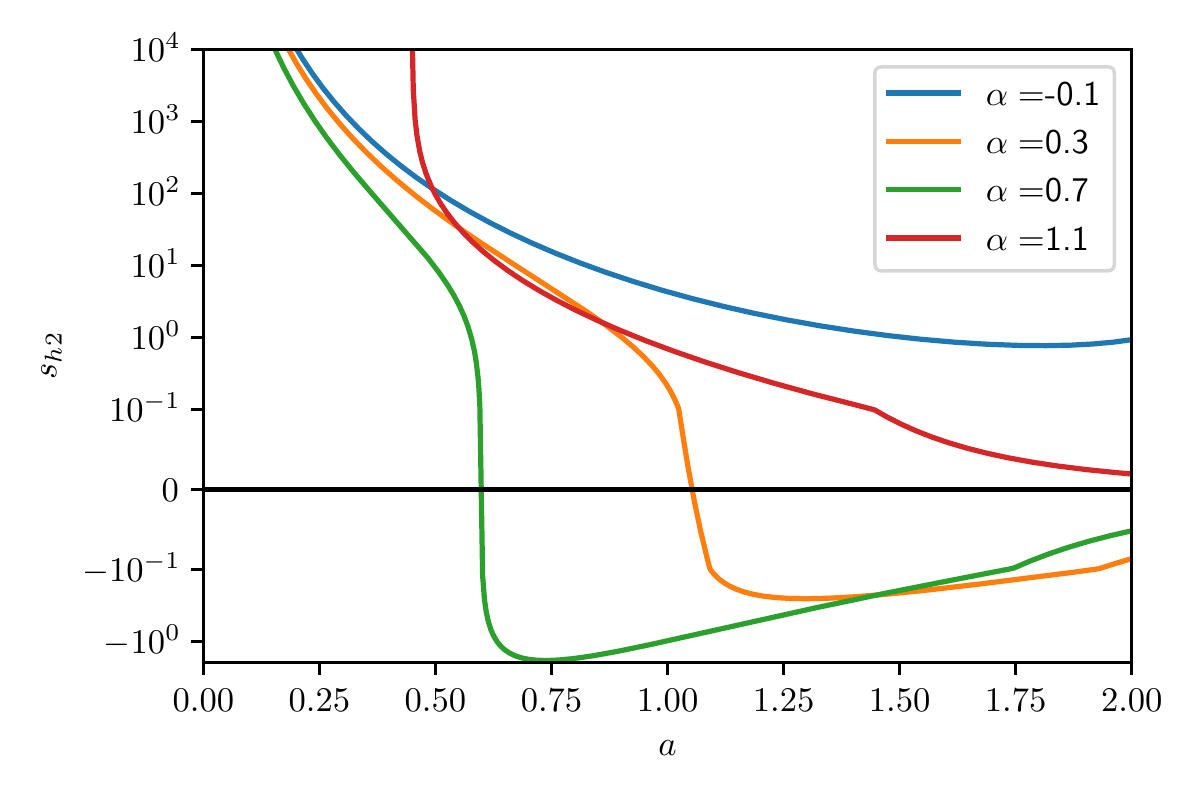}
 % M1s1.pdf: 0x0 px, 300dpi, 0.00x0.00 cm, bb=
 \caption{Model $M_1$: $s_1$ and $s_{h2}$ as function of scale factor for some values of $\alpha$. We have used a mixed log-linear scale in order to view both asymptotic behaviour and zero crossing.}
 \label{m1s}
\end{figure}

In Figure \ref{m1s}, we may see that $s_1\geq0$ for $\alpha\leq1$ and $s_{h2}(a\rightarrow\infty)<0$ for $0<\alpha<1$, in agreement with our analysis. As discussed above, $s_{h2}<0$ implies $s_{m2}<0$, so we choose to plot only $s_1$ and $s_{h2}$ for each model, for clarity.

\subsection{\texorpdfstring{$M_2: \Gamma=3\alpha H_0$}{M2: Gamma = 3 alpha H0}}
In this case we have the fixed parameter values $\beta=0$ and $n=0$, so from (\ref{Slingt0}) we see that $S'\geq0$ reads 
\be
s_1=3(1-\alpha)a^{-\frac{3}{2}}\geq0.
%\label{Slingt0}
\ee
which implies $\alpha\leq1$. From (\ref{expSh}), the condition $S''_h<0$ reads
\be
s_{h2}=\frac{3}{4}\left(1-\alpha\right) a^{-\frac{7}{2}}\left[4(1-\alpha)a^{-\frac{3}{2}}-5\alpha\right]<0
%\label{expSh}
\ee
Thus, for $a\rightarrow\infty$, it implies $0<\alpha<1$. From (\ref{expSm}), the condition $S''_m<0$ reads
\be
s_{m2}=\frac{3}{4}\left(1-\alpha\right)a^{-\frac{7}{2}}\left[(1-\alpha)a^{-\frac{3}{2}}-5\alpha\right]<0
%\label{expSm}
\ee
which yields the same limit for $a\rightarrow\infty$, $0<\alpha<1$.

\begin{figure}[ht]
 \centering
 \includegraphics[width=.48\textwidth]{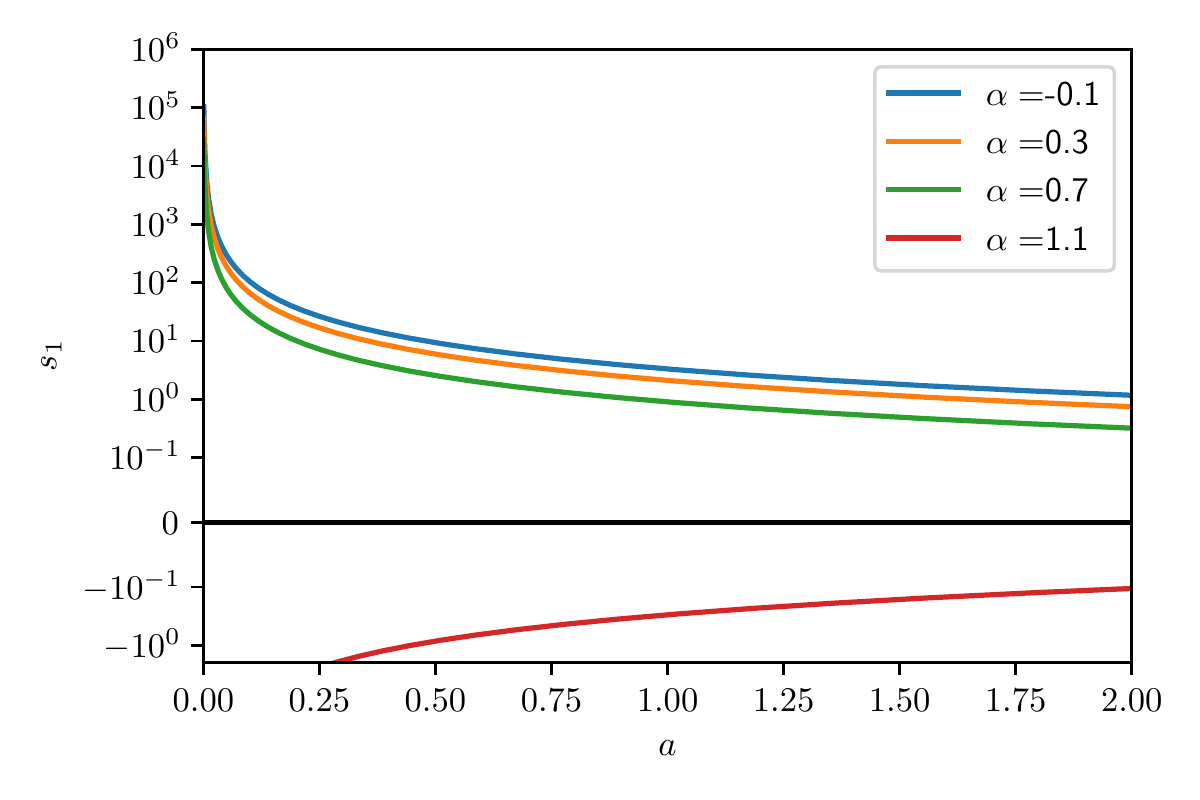}
 \includegraphics[width=.48\textwidth]{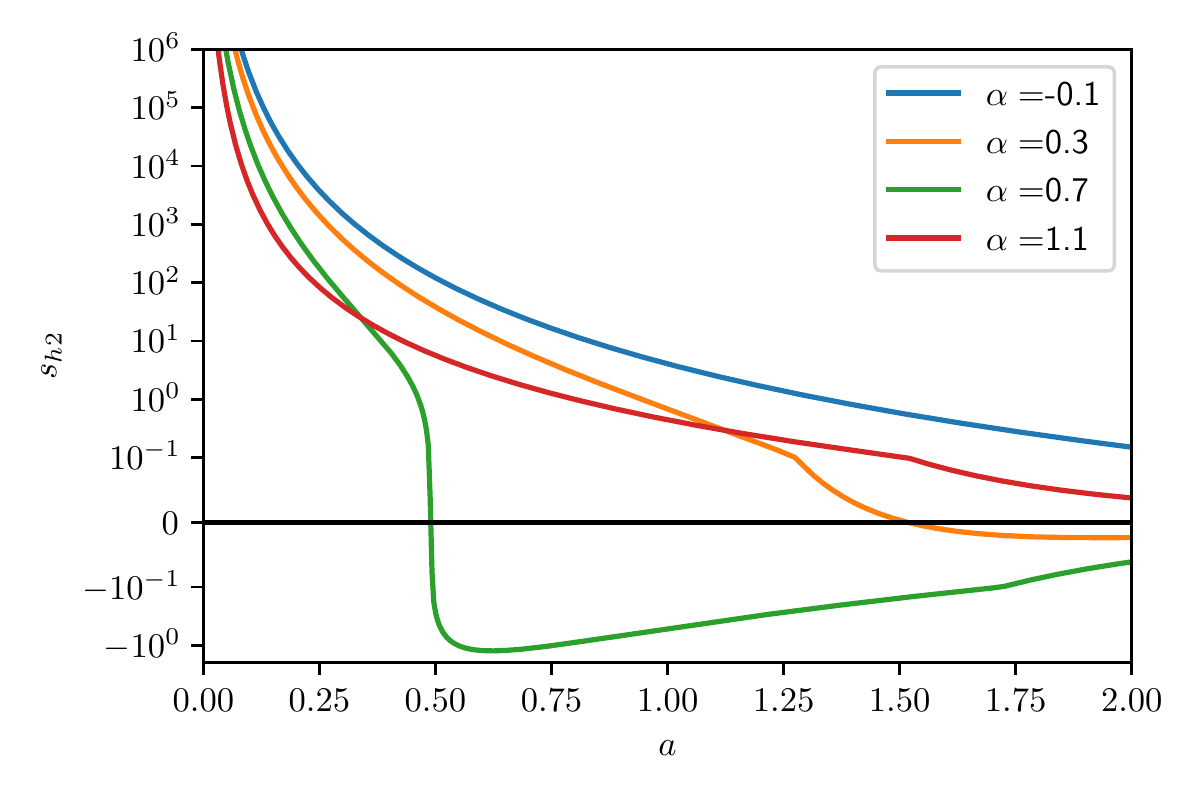}
 % M1s1.pdf: 0x0 px, 300dpi, 0.00x0.00 cm, bb=
 \caption{Model $M_2$: $s_1$ and $s_{h2}$ as function of scale factor for some values of $\alpha$. We have used a mixed log-linear scale in order to view both asymptotic behaviour and zero crossing.}
 \label{m2s}
\end{figure}

In Figure \ref{m2s}, we may see that $s_1\geq0$ for $\alpha\leq1$ and $s_{h2}(a\rightarrow\infty)<0$ for $0<\alpha<1$, in agreement with our analysis.

\subsection{\texorpdfstring{$M_3$: $\Gamma=3\beta H$}{M3: Gamma = 3 beta H}}
In this case we have the fixed parameter value $\alpha=0$, so from (\ref{Slingt0}) we see that $S'\geq0$ reads 
\be
s_1=(1-\beta) a^{\frac{3}{2}(\beta -1)}\geq0.
%\label{Slingt0}
\ee
which implies $\beta\leq1$. From (\ref{expSh}), the condition $S''_h<0$ reads
\be
s_{h2}=\frac{3}{2}(1-\beta)(2-3\beta)a^{3\beta-5}<0
%\label{expSh}
\ee
Thus, it implies $\frac{2}{3}<\beta<1$. From (\ref{expSm}), the condition $S''_m<0$ reads
\be
s_{m2}=\frac{3}{4}(1-\beta) (1-3\beta) a^{3 \beta -5}<0
%\label{expSm}
\ee
which yields the limit $\frac{1}{3}<\beta<1$. As one may see, for all the interval that we have $S''_h<0$ we have also $S''_m<0$, as expected.

\begin{figure}[ht]
 \centering
 \includegraphics[width=.48\textwidth]{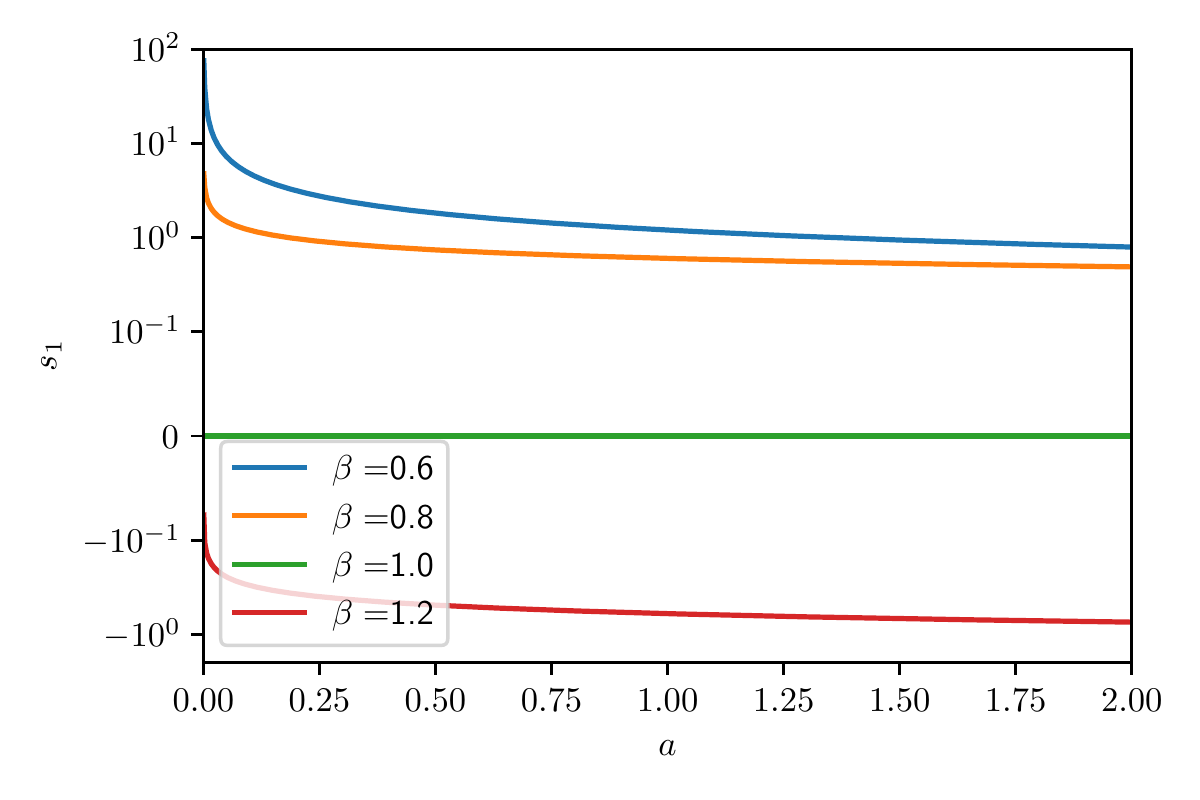}
 \includegraphics[width=.48\textwidth]{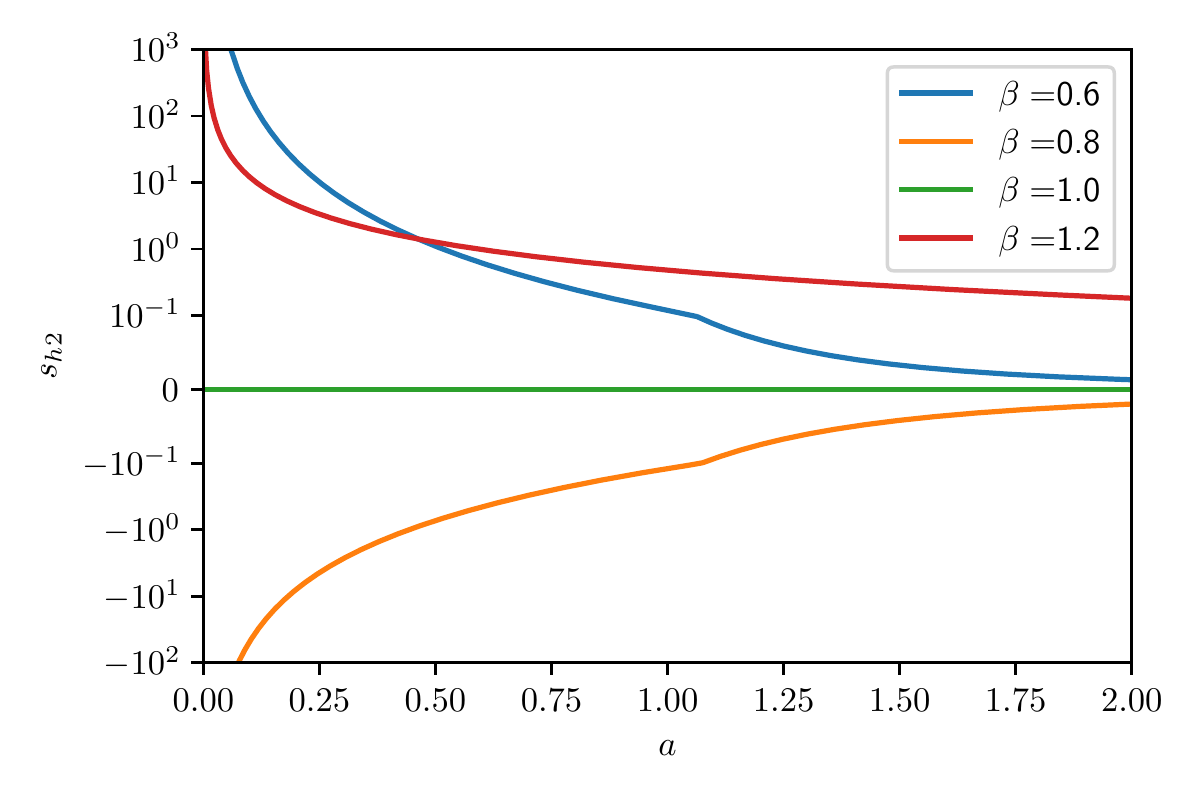}
 % M1s1.pdf: 0x0 px, 300dpi, 0.00x0.00 cm, bb=
 \caption{Model $M_3$: $s_1$ and $s_{h2}$ as function of scale factor for some values of $\beta$. We have used a mixed log-linear scale in order to view both asymptotic behaviour and zero crossing.}
 \label{m3s}
\end{figure}

In Figure \ref{m3s}, we may see that $s_1\geq0$ for $\beta\leq1$ and $s_{h2}(a\rightarrow\infty)<0$ for $\frac{2}{3}<\beta<1$, in agreement with our analysis.

\subsection{\texorpdfstring{$M_4: \Gamma=3\alpha H_0\left(\frac{H_0}{H}\right)^n$}{M4: Gamma = 3 alpha H0 (H0/H)**n}}
In this case we have the fixed parameter value $\beta=0$, so from (\ref{Slingt0}) we see that $S'\geq0$ reads 
\be
s_1=3(1-\alpha)\left(\frac{H}{H_0}\right)^{-n}a^{-\frac{3}{2}(n+1)}\geq0.
%\label{Slingt0}
\ee
which implies $\alpha\leq1$. From (\ref{expSh}), the condition $S''_h<0$ reads
\be
s_{h2}=\frac{3}{4}\left(\frac{H}{H_0}\right)^{-2n}\left(1-\alpha\right)a^{-2-\frac{3}{2}(n+1)}\left[4(1-\alpha)a^{-\frac{3}{2}(n+1)}-\alpha\left(3n+5\right)\right]<0
%\label{expSh}
\ee
For $a\rightarrow\infty$, there are some subcases here, according to the sign of the exponent $-\frac{3}{2}(n+1)$, that is, if $n$ is greater than $-1$ or not. If $n>-1$, the condition can be summarized as $\alpha(\alpha-1)(3n+5)<0$. As $3n+5>0$, it implies $0<\alpha<1$. If $n<-1$, the condition is $(\alpha-1)^2<0$, which is impossible, so $n<-1$ is discarded by this analysis. In the special case of $n=-1$, we recover the model $M_3$, so $\frac{2}{3}<\alpha<1$.

From (\ref{expSm}), the condition $S''_m<0$ reads
\be
s_{m2}=\frac{3}{4}\left(\frac{H}{H_0}\right)^{-2n}\left(1-\alpha\right)a^{-2-\frac{3}{2}(n+1)}\left[(1-\alpha)a^{-\frac{3}{2}(1+n)}-\alpha\left(3n+5\right)
\right]<0
%\label{expSm}
\ee
which yields the same limit for $a\rightarrow\infty$ and $n>-1$: $0<\alpha<1$. Just like before, $n=-1$ implies, like in $M_3$, $\frac{1}{3}<\alpha<1$.

\begin{figure}[ht]
 \centering
\includegraphics[width=.48\textwidth]{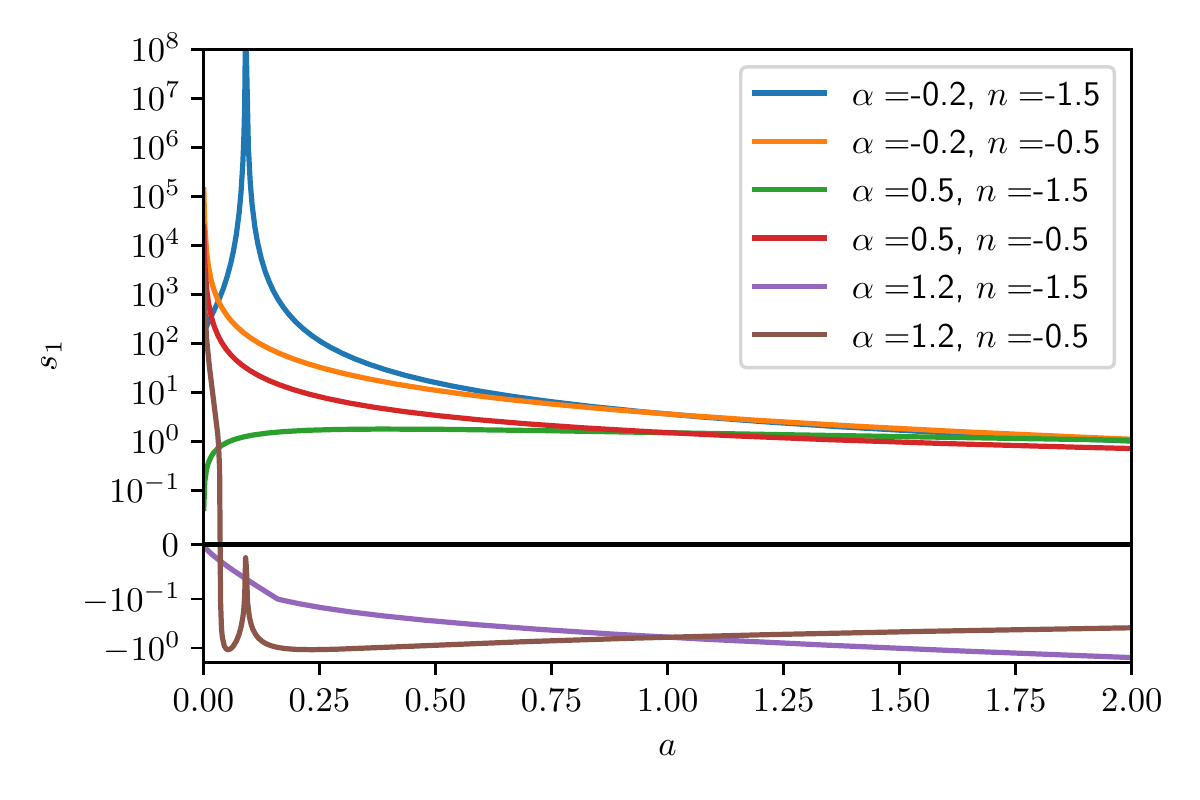}
\includegraphics[width=.48\textwidth]{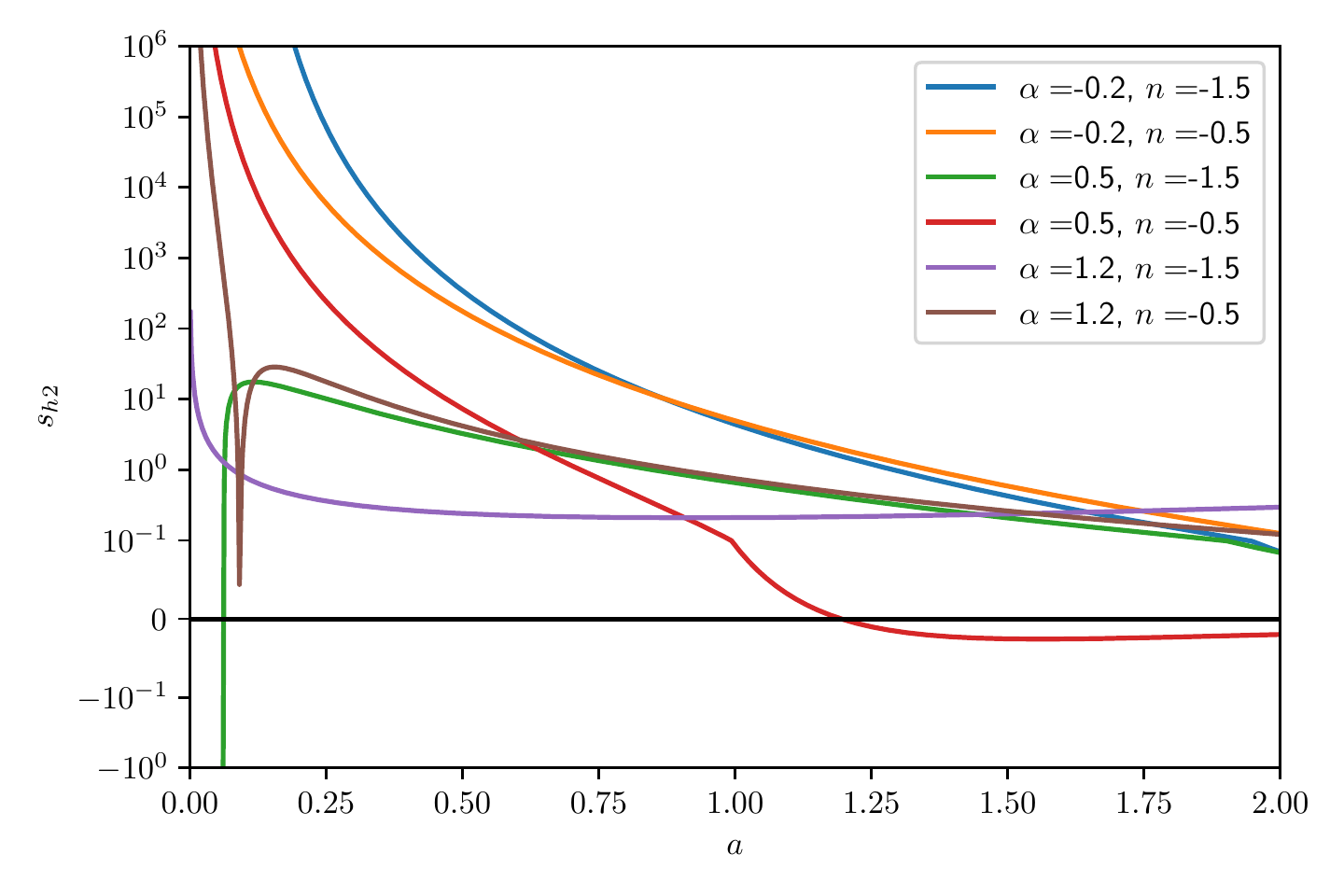}
% \includegraphics[width=.48\textwidth]{fig/M4s1.pdf}
 % M1s1.pdf: 0x0 px, 300dpi, 0.00x0.00 cm, bb=
 \caption{Model $M_4$: $s_1$ and $s_{h2}$ as function of scale factor for some values of $(\alpha,n)$. We have used a mixed log-linear scale in order to view both asymptotic behaviour and zero crossing.}
 \label{m4s}
\end{figure}

In Figure \ref{m4s}, we may see that $s_1\geq0$ for $\alpha\leq1$ and $s_{h2}(a\rightarrow\infty)<0$ for $0<\alpha<1$ and $n>-1$, in agreement with our analysis.

\subsection{\texorpdfstring{$M_5: \Gamma=3\alpha\frac{H_0^2}{H}+3\beta H$}{M5: Gamma = 3 alpha H0**2/H + 3 beta H}}
%\subsection{\texorpdfstring{$M_5: $}{M5: }}
In this case we have the fixed parameter value $n=1$, so from (\ref{Slingt0}) we see that $S'\geq0$ reads 
\be
s_1=3(1-\alpha-\beta)\left(\frac{H_0}{H}\right)a^{-3(1-\beta)}\geq0.
%\label{Slingt0}
\ee
which implies $1-\alpha-\beta\geq0$. From (\ref{expSh}), the condition $S''_h<0$ reads
\be
s_{h2}=\frac{3}{2}\left(\frac{H_0}{H}\right)^2\left(\frac{1-\alpha-\beta}{1-\beta}\right)a^{-5+3\beta}\left[(1-\alpha-\beta)(2-3\beta)a^{-3(1-\beta)}+\alpha\left(3\beta-4\right)\right]<0
%\label{expSh}
\ee

To analyze the behaviour for $a\rightarrow\infty$ we have to make assumptions about the scale factor exponent, $-3(1-\beta)$. If $\beta<1$, $S''_h<0$ implies $\alpha(1-\alpha-\beta)(3\beta-4)<0$. If we combine with the condition from $s_1$, we must have $1-\alpha-\beta>0$, thus it simplifies to $\alpha(3\beta-4)<0$. Thus, $\alpha>0$ and $\beta<\frac{4}{3}$ or $\alpha<0$ and $\beta>\frac{4}{3}$.

For $\beta>1$, $S''_h<0$ would imply $\beta<\frac{2}{3}$, so $\beta>1$ is not allowed by this analysis.

If $\beta=1$, Eq. (\ref{Habeta1}) with $n=1$ yields
\be
E=\left[1+3\alpha\ln a\right]^{1/2},
\label{Habetan1}
\ee
from which we find
\be
s_{h2}=\frac{3\alpha}{2a^2}\left(\frac{H_0}{H}\right)^2(1+6\alpha+3\alpha\ln a)
\ee

In this case, in the limit $a\rightarrow\infty$, $S''_h<0$ implies $\alpha^2<0$, that is, $\beta=1$ is not allowed by this analysis.

From (\ref{expSm}), the condition $S''_m<0$ reads
\be
s_{m2}=\frac{3}{4}\left(\frac{H_0}{H}\right)^2\left(\frac{1-\alpha-\beta}{1-\beta}\right)a^{-5+3\beta}\left[(1-\alpha-\beta)(1-3\beta)a^{-3(1-\beta)}+2\alpha\left(3\beta-4\right)\right]<0
%\label{expSm}
\ee

In this case, in the limit $a\rightarrow\infty$, for $\beta<1$, $S''_m<0$ implies $\alpha(3\beta-4)(1-\alpha-\beta)<0$. Combining it with the condition from $s_1$, we have $1-\alpha-\beta>0$, thus $\alpha(3\beta-4)<0$. So, if $\alpha>0$, $\beta<\frac{4}{3}$ and if $\alpha<0$, we have $\beta>\frac{4}{3}$.

For $\beta>1$, $S''_m<0$ would imply $\beta<\frac{1}{3}$, so $\beta>1$ is not allowed by this analysis.

\begin{figure}[ht]
 \centering
 \includegraphics[width=.8\textwidth]{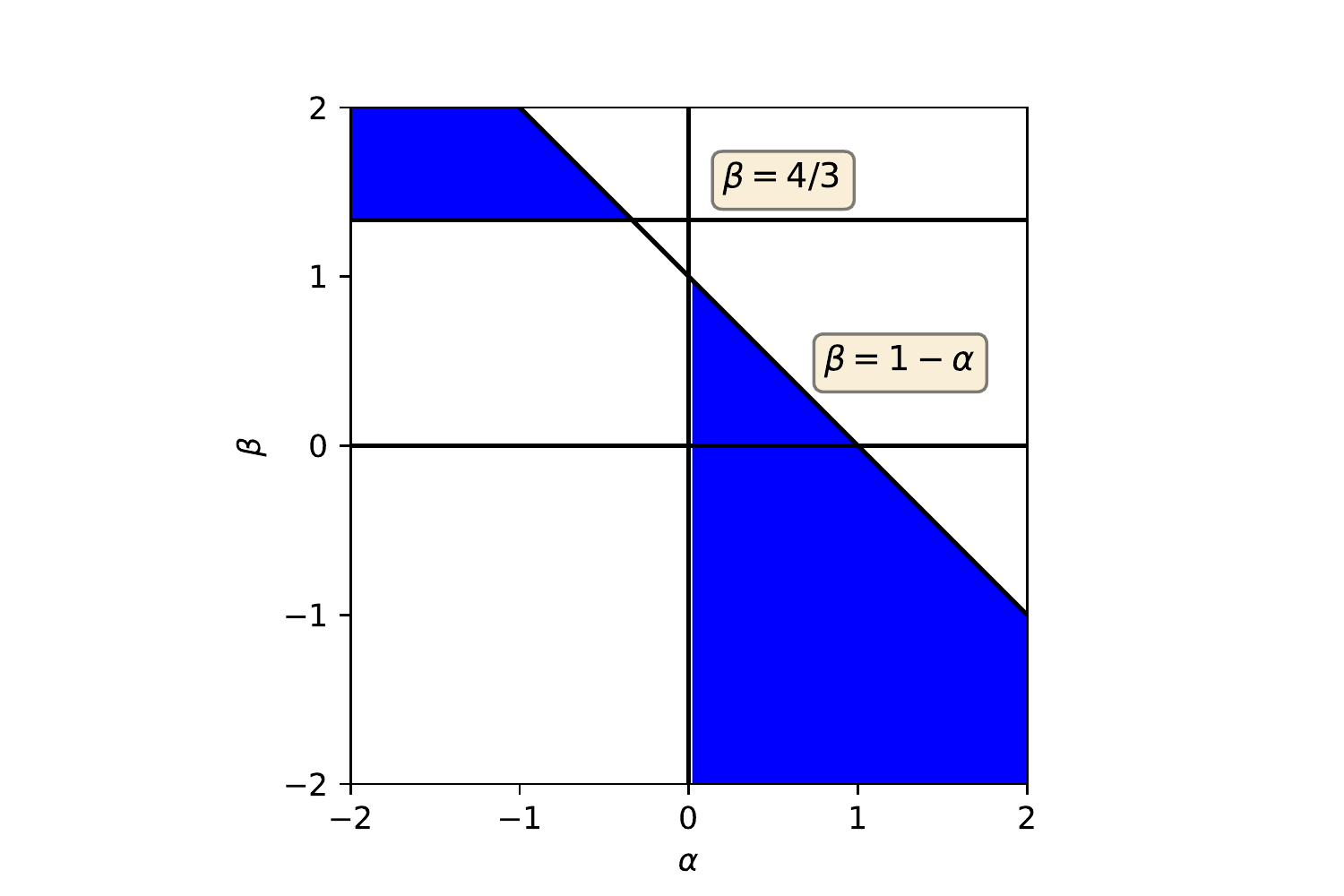}
 % M5limits.pdf: 0x0 px, 300dpi, 0.00x0.00 cm, bb=
 \caption{Limits over free parameters for model $M_5$. The blue regions correspond to values of parameters allowed by the conditions $S'\geq0$, $S''_h<0$ and $S''_m<0$.}
 \label{M5limits}
\end{figure}

For $\beta=1$, $s_{m2}$ is written:
\be
s_{m2}=\frac{3\alpha}{4a^2}\left(\frac{H_0}{H}\right)^2(2+9\alpha+6\alpha\ln a)
\ee
In this case, in the limit $a\rightarrow\infty$, $S''_m<0$ implies $\alpha^2<0$, that is, $\beta=1$ is not allowed by this analysis. The limits for model $M_5$ can be viewed on Fig. \ref{M5limits}.

\begin{figure}[!ht]
 \centering
 \includegraphics[width=.49\textwidth]{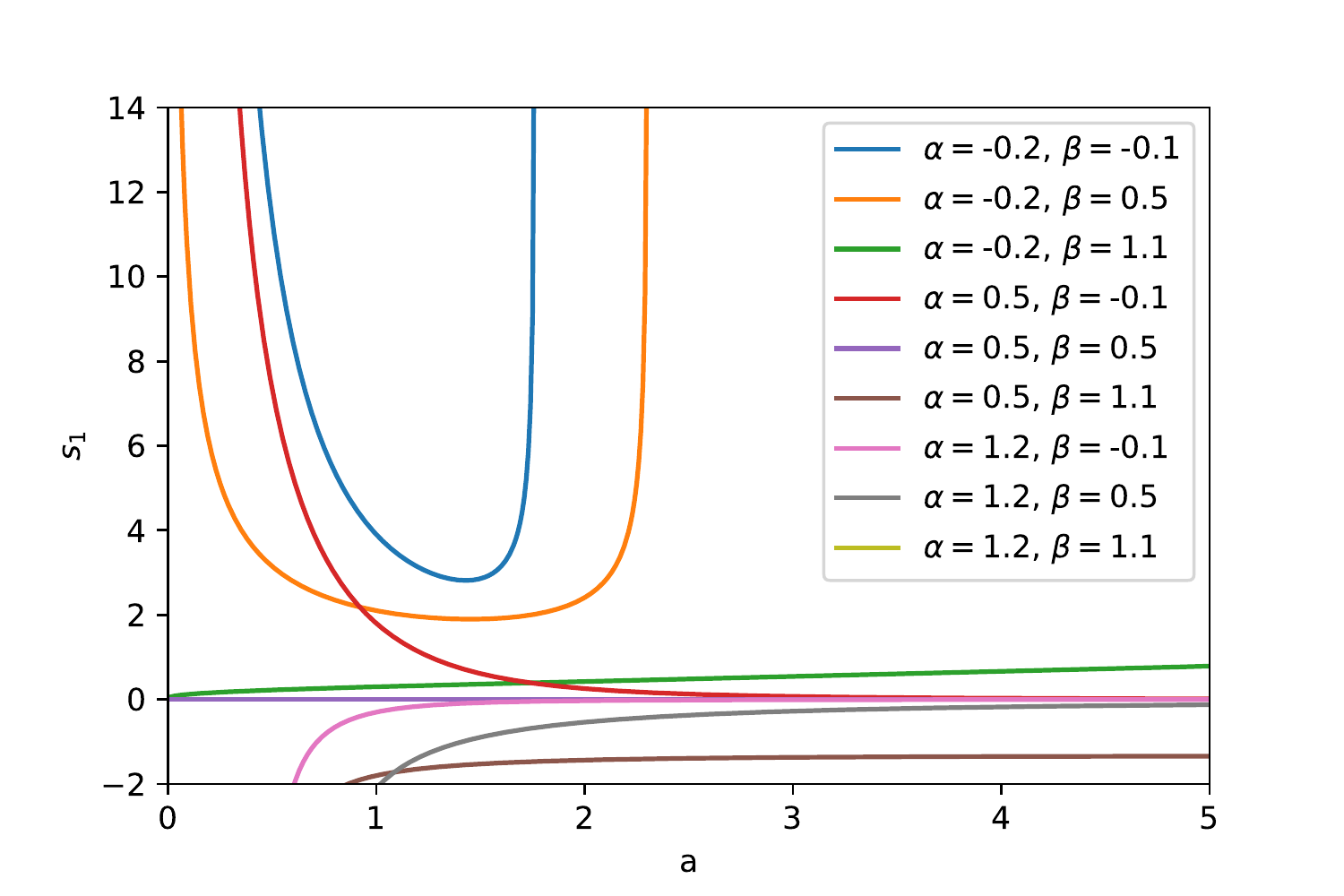}
 \includegraphics[width=.49\textwidth]{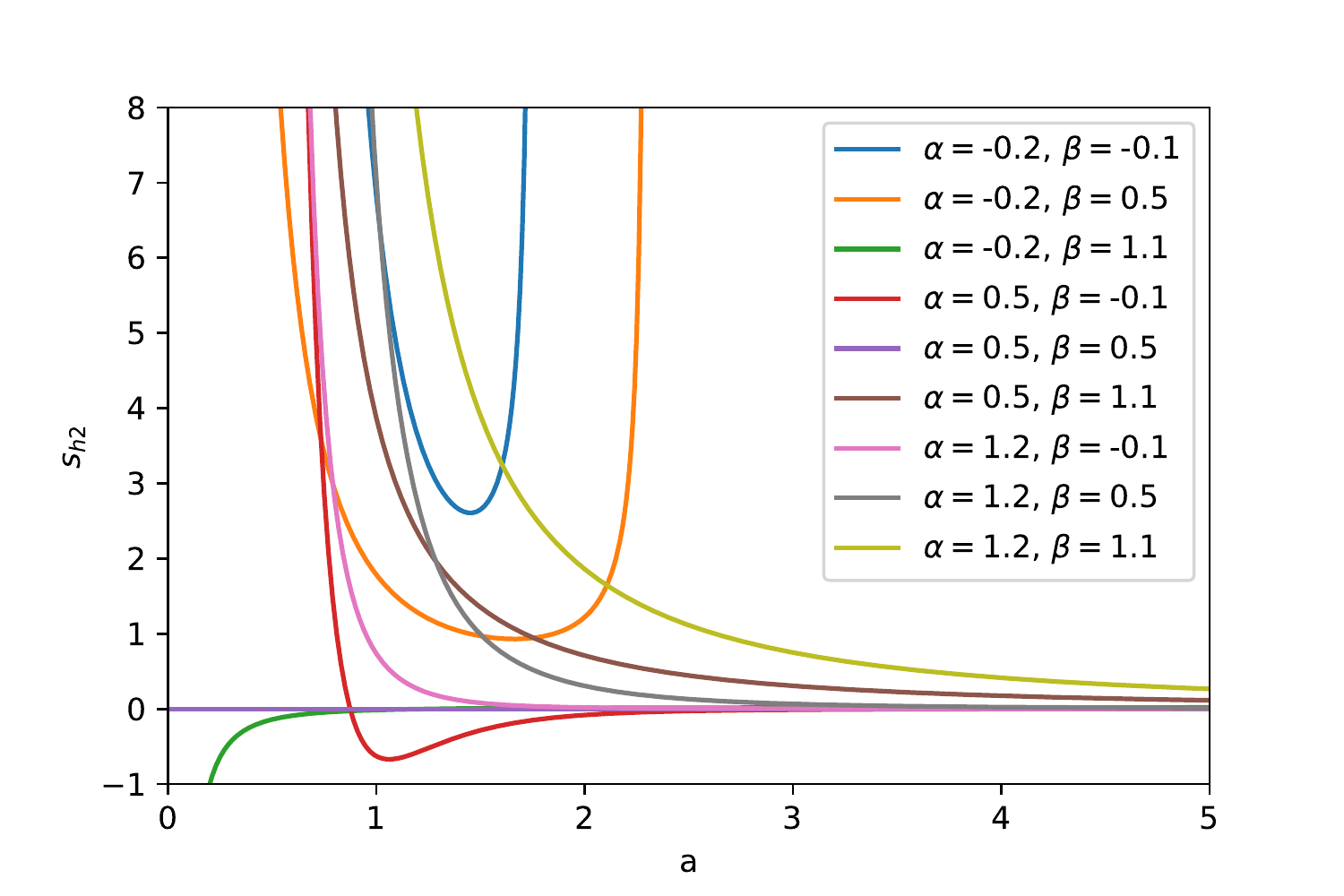}
 % M1s1.pdf: 0x0 px, 300dpi, 0.00x0.00 cm, bb=
 \caption{Model $M_5$: $s_1$ and $s_{h2}$ as function of scale factor for some values of $(\alpha,\beta)$.}
 \label{m5s}
\end{figure}

In Figure \ref{m5s}, we may see that $s_1\geq0$ for $\alpha\leq1$ and $s_{h2}(a\rightarrow\infty)<0$ for $0<\alpha<1$ and $n>-1$, in agreement with our analysis.

The results of all models from Table \ref{models} can be seen on Table \ref{tab2}.

\begin{table}[ht]
\renewcommand{\arraystretch}{1.2} % aumenta line spacing
	\centering
		\begin{tabular}{|c|c|c|c|c|c|}
\hline\hline
Model &  Creation rate & $S'\geq0$ & $S''_h<0$ & $S''_m<0$ & Combination\\
\hline
$M_1$	& $\Gamma=\frac{3\alpha H_0^2}{H}$ & $\alpha\leq1$ & $0<\alpha<1$ & $0<\alpha<1$ & $0<\alpha<1$\\
$M_2$	& $\Gamma=3\alpha H_0$ & $\alpha\leq1$ & $0<\alpha<1$ & $0<\alpha<1$ & $0<\alpha<1$\\
$M_3$	& $\Gamma=3\beta H$ & $\beta\leq1$ & $\frac{2}{3}<\beta<1$ & $\frac{1}{3}<\beta<1$ & $\frac{2}{3}<\beta<1$\\
$M_4$	& $\Gamma=3\alpha H_0\left(\frac{H_0}{H}\right)^n$ & $\alpha\leq1$ & $0<\alpha<1$, $n>-1$ & $0<\alpha<1$, $n>-1$ & $0<\alpha<1$, $n>-1$\\
$M_5$	& $\Gamma=3\alpha\frac{H_0^2}{H}+3\beta H$ & $1-\alpha-\beta\geq0$ & $\alpha(3\beta-4)<0$ & $\alpha(3\beta-4)<0$ & $\alpha>0$, $\beta\leq1-\alpha$\\
\hline
\hline
		\end{tabular}\\
\caption{Thermodynamic constraints on free parameters of matter creation models.}
\label{tab2}
\end{table}

\section{\label{discussion} Discussion and Concluding Remarks}
We have analyzed the thermodynamics of 5 spatially flat CCDM models, taking into account a contribution from the horizon entropy, based on Holographic Principle.

In principle, the initial state of de Sitter age should be stable ($H$ and $S$ constants when $t\rightarrow\infty$) but particle creation ($\Gamma$), according to \cite{MimosoPavon2013}, can be seen as an external agent acting on the system. Before the thermodynamic equilibrium was reached, the Universe needed to self-adjust to allow the ultimate expansion of de Sitter through the ages of radiation and matter.

The rate of particle production is irreversible, in this case for the five models treated in this work. In practice, irreversibility directly implies the generation of entropy \cite{Calvao1992}, as well as the increase in volume in the phase space. In our analysis, the particle production rate $\Gamma$ for the five models analyzed, was implicitly or explicitly included in the expressions for $S'$ and $ S''$, as can be seen in equations (\ref{dSda}) and (\ref{d2Sda2}). For easy of analysis we defined the quantities $s_1$ for the first derivative and $s_{h2}$ and $s_{m2}$ for the second order derivatives. All models discussed in this work are particular cases of the general model with three free parameters: $\alpha$, $\beta$ and $n$. The $M_{1}$ model has only one free parameter $\alpha$ and the analysis of the derivatives suggests that it is between $ 0<\alpha<1$. $M_{2}$ is a model similar to $M_{1}$ but with constant $\Gamma$, the limits for $\alpha$ is $0<\alpha<1$. The $M_{3}$ model has $\beta$ as a free parameter and $\Gamma$ varies linearly with $H$ and $\frac{2}{3}<\beta<1$. $M_{4}$ has two free parameters: $\alpha$ and $n$, $\Gamma$ is a power law over $H$: $\Gamma\propto H^{-n}$. The validity interval was $0<\alpha <1$ with $ n> - 1$, for $n = 0$ $M_{4}$ corresponds to $M_{2}$ and if $n = 1$ it becomes $M_{1}$. For the model $M_{5}$, which is a combination of $M_{1}$ and $M_{3}$, $\beta\leq 1-\alpha$ and $\alpha(3\beta-4)<0$.

The limits over the parameters $\alpha$ and $\beta$ could be seen in figure \ref{M5limits}.

It is also interesting to mention that some of the models analyzed here can lead to singularities in the future ($H\rightarrow0$) and to see how it compares with the thermodynamic constraints we found. Models $M_1$--$M_3$ give no singularity at the future. Models $M_4$ and $M_5$ yield future singularities for some regions of the parameters. Model $M_4$ will have future singularity for $n>-1$ and $\alpha<0$ or $n<-1$ and $\alpha>1$. It is important to notice that this region is disallowed from our thermodynamic analysis. Model $M_5$ will have future singularity for $\beta<1$ and $\alpha<0$ or $\beta>1$ and $\alpha<1-\beta$. For this model, the thermodynamic analysis allows a future singularity only for $\beta>4/3$ and $\alpha<1-\beta$. Concerning our thermodynamic analysis, we found no problem with this region of the parameter space.

Further analysis of matter creation models may include the conserved baryonic contribution and spatial curvature. It could be interesting to test if the baryonic contribution, although small, could give non-negligible changes to the constraints we found. It is also interesting to see if this thermodynamic analysis could contribute to the current tension of constraints over the spatial curvature \cite{DiValentino2019}. Other creation rates not considered here could also be analyzed.

\begin{acknowledgments}
JFJ has been supported by  {\it Funda\c{c}\~ao de Amparo \`a Pesquisa do Estado de S\~ao Paulo} - FAPESP (Process no. 2017/05859-0) and  R. V. has been supported by  {\it Funda\c{c}\~ao de Amparo \`a Pesquisa do Estado de S\~ao Paulo} - FAPESP (Process no. 2013/26258-4 and 2016/09831-0). This study was financed in part by the Coordena\c{c}\~ao de Aperfei\c{c}oamento de Pessoal de N\'ivel Superior - Brasil (CAPES) - Finance Code 001.
\end{acknowledgments}


\begin{thebibliography}{99}
\bibitem{Callen1985} H.~B.~Callen, ``Thermodynamics and an Introduction to Thermostatistics'', 2nd Edition, pp.~512.%~ISBN 0-471-86256-8.
~Wiley-VCH,% August
 1985.

\bibitem{MimosoPavon2013} J.~P.~Mimoso \& D.~Pav{\'o}n, \prd, 87, 047302, 2013. 

\bibitem{t'Hooft1993} G.~'t Hooft, 1993, arXiv:gr-qc/9310026. 

\bibitem{Susskind1995} L.~ Susskind, Journal of Mathematical Physics, 36, 6377, 1995.

\bibitem{FischlerSusskind1998} W.~Fischler \& L.~Susskind, arXiv:hep-th/9806039, 1998.

\bibitem{BakRey2000} D.~Bak \& S.~J.~Rey, Classical and Quantum Gravity, 17, L83, 2000.

\bibitem{Parker1968} 
L.~ Parker,
Phys.\ Rev.\ Letters, {\bf vol. 21}, 8, pp. 562-564, 1968.

\bibitem{PrigogineEtAl1988}
I.~ Prigogine, J.~ Geheniau, E.~ Gunzig, P.~ Nardone, 1989, Gen. Relativ. Grav.,
21, 767, 1989.

\bibitem{PanEtAl2016}
S.~ Pan, J.~ de Haro, A.~ Paliathanasis and R.~ J.~ Slagter, Monthly Notices of Royal Astronomy Society,  460, 1445–1456 (2016).

\bibitem{HagiwaraEtAl2002}
K.~ Hagiwara et al., Phys. Rev. D, 66, 010001, 2002.

\bibitem{PeeblesRatra2003}
P.~ J.~ E.~ Peebles, B.~ Ratra, Rev. Mod. Phys., 75, 559, 2002.

\bibitem{EllisEtAl1989}
J.~ Ellis, S.~ Kalara, K.~A.~ Olive, C.~ Wetterich, Phys. Lett. B, 228, 264, 1989.


\bibitem{SteigmanEtAl2009}
G.~ Steigman, R.~C.~ Santos, J.~A.~S.~ Lima, J. Cosmol. Astropart. Phys.,
06, 033, 2009.

\bibitem{FabrisEtAl2014}
J.~ C.~ Fabris ,J.~A.~F.~ Pacheco, O.~F.~ Piattella, J. Cosmol. Astropart.
Phys., 06, 038, 2014.

\bibitem{LimaEtAl2014}
J.~ A.~ S.~ Lima, L.~L.~Graef, D.~Pav\'on, S.~Basilakos, J. Cosmol. As-
tropart. Phys., 10, 042, 2014.

\bibitem{ChakrabortyEtAl2015}
S.~ Chakraborty, S.~ Pan, S.~ Saha, 2015, preprint (arXiv:1503.05552)


\bibitem{LimaEtAl2010}
J.~A.~S.~ Lima, J.~F.~ Jesus,  F.~A.~ Oliveira, J. Cosmol. Astropart. Phys.,
11, 027, 2010.

\bibitem{NunesPavon2015}
R.~ C.~ Nunes, D.~ Pav\'on, 2015, Phys. Rev. D, 91, 063526.

\bibitem{Cadwell2002}
R.~ R.~ Caldwell, Physics Letters B, 545, 23-29, 2002. 

\bibitem{PanChakraborty2015}
S.~ Pan, S.~ Chakraborty, Adv. High Energy Phys., 654025, 2015.

\bibitem{ChakrabortyEtAl2014}
S.~Chakraborty, S.~Pan, S.~Saha, Phys. Lett. B, 738, 424, 2014.

\bibitem{PanEtAl2019}
S.~Pan, J.~D.~Barrow, A.~ Paliathanasis, Eur. Phys. J. C, 79, 115, 2019.

%\cite{Jesus:2016lte}
\bibitem{bic-ccdm}
  J.~F.~Jesus, R.~Valentim and F.~Andrade-Oliveira,
  %``Bayesian analysis of CCDM Models,''
  JCAP {\bf 1709} (2017) no.09,  030
  %doi:10.1088/1475-7516/2017/09/030
  [arXiv:1612.04077 [astro-ph.CO].
  %%CITATION = doi:10.1088/1475-7516/2017/09/030;%%

 \bibitem{DiValentino2019}
       E.~Di Valentino, A.~Melchiorri and J.~Silk,
       % title = "{Planck evidence for a closed Universe and a possible crisis for cosmology}",
      Nature Astronomy, (2020), 196-203
   %  keywords = {Astrophysics - Cosmology and Nongalactic Astrophysics, General Relativity and Quantum Cosmology, High Energy Physics - Phenomenology, High Energy Physics - Theory},
         [arXiv:1911.02087 [astro-ph.CO]]

\bibitem{GraefEtAl2014}
L.~L.~ Graef, D.~ Pav\'on., S.~ Basilakos, J. Cosmol. Astropart. Phys., 10, 042, 2014.


\bibitem{LimaBasilakos2012} J.~A.~S.~Lima, S.~Basilakos, \& F.~E.~M.~Costa, \prd, 86, 103534, 2012. 

%\bibitem{LimaBasilakos2013a} Lima, J.~A.~S., Basilakos, S., \& Sol{\`a}, J.\ 2013, \mnras, 431, 923.

\bibitem{LimaBasilakos2013} J.~A.~S.~Lima, S.~Basilakos, \& J.~ Sol{\`a}, \mnras, 431, 923, 2013.

\bibitem{Freaza2002} M.~P.~ Freaza, R.~S.~de Souza, \& I.~Waga, \prd, 66, 103502, 2002.

\bibitem{Lima2010} J.~A.~S.~Lima, J.~F.~Jesus, \&F.~A.~Oliveira, \jcap, 11, 027, 2010. 

\bibitem{Lima2008} J.~A.~S.~ Lima, F.~E.~Silva, \& R.~C.~Santos, Classical and Quantum Gravity, 25, 205006, 2008.

%\bibitem{Jesus2017} Jesus, J.~F., Valentim, R. \& Andrade-Oliveira, F.\ 2017, \jcap, 1, 030.

%\cite{Jesus:2015wua}
\bibitem{JO16} 
  J.~F.~Jesus and F.~Andrade-Oliveira,
  %``CCDM Model with Spatial Curvature and The Breaking of "Dark Degeneracy",''
  JCAP {\bf 1601}, 014 (2016)
%  doi:10.1088/1475-7516/2016/01/014
  [arXiv:1503.02595 [astro-ph.CO]].
  %%CITATION = doi:10.1088/1475-7516/2016/01/014;%%
  %1 citations counted in INSPIRE as of 13 Oct 2016

%\cite{Jesus:2014gha}
\bibitem{JesusPereira14}
  J.~F.~Jesus and S.~H.~Pereira,
  %``CCDM model from quantum particle creation: constraints on dark matter mass,''
  JCAP {\bf 1407} (2014) 040
  %doi:10.1088/1475-7516/2014/07/040
  [arXiv:1403.3679 [astro-ph.CO]].
  %%CITATION = doi:10.1088/1475-7516/2014/07/040;%%
  %19 citations counted in INSPIRE as of 09 Feb 2020
  
%\cite{Lima:2014hda}
\bibitem{LimaBaranov14}
  J.~A.~S.~Lima and I.~Baranov,
  %``Gravitationally Induced Particle Production: Thermodynamics and Kinetic Theory,''
  Phys.\ Rev.\ D {\bf 90} (2014) no.4,  043515
  %doi:10.1103/PhysRevD.90.043515
  [arXiv:1411.6589 [gr-qc]].
  %%CITATION = doi:10.1103/PhysRevD.90.043515;%%
  %37 citations counted in INSPIRE as of 09 Feb 2020

%\cite{Lima:2009ic}
\bibitem{ljo10} 
  J.~A.~S.~Lima, J.~F.~Jesus and F.~A.~Oliveira,
  %``CDM Accelerating Cosmology as an Alternative to LCDM model,''
  JCAP {\bf 1011}, 027 (2010)
  [arXiv:0911.5727 [astro-ph.CO]].
  %%CITATION = ARXIV:0911.5727;%%
  %28 citations counted in INSPIRE as of 27 Feb 2015

%\cite{Radicella:2010zb}
\bibitem{RadPav12}
  N.~Radicella and D.~Pavon,
  %``A thermodynamic motivation for dark energy,''
  Gen.\ Rel.\ Grav.\  {\bf 44} (2012) 685
  %doi:10.1007/s10714-011-1299-y
  [arXiv:1012.0474 [gr-qc]].
  %%CITATION = doi:10.1007/s10714-011-1299-y;%%
  %54 citations counted in INSPIRE as of 21 Apr 2019

%\cite{Graef:2013iia}
\bibitem{GraefEtAl14} 
  L.~L.~Graef, F.~E.~M.~Costa and J.~A.~S.~Lima,
  %``On the equivalence of $\Lambda(t)$ and gravitationally induced particle production cosmologies,''
  Phys.\ Lett.\ B {\bf 728}, 400 (2014)
  [arXiv:1303.2075 [astro-ph.CO]].
  %%CITATION = ARXIV:1303.2075;%%
  %5 citations counted in INSPIRE as of 18 Sep 2014
  
%\cite{Calvao:1991wg}
\bibitem{Calvao1992}
  M.~O.~Calv\~ao, J.~A.~S.~Lima and I.~Waga,
  %``On the thermodynamics of matter creation in cosmology,''
  Phys.\ Lett.\ A {\bf 162} (1992) 223.
  %doi:10.1016/0375-9601(92)90437-Q
  %%CITATION = doi:10.1016/0375-9601(92)90437-Q;%%
  %173 citations counted in INSPIRE as of 17 Apr 2019
  
%\cite{Lima:2012cm}
\bibitem{Lima2012}
  J.~A.~S.~Lima, S.~Basilakos and F.~E.~M.~Costa,
  %``New Cosmic Accelerating Scenario without Dark Energy,''
  Phys.\ Rev.\ D {\bf 86} (2012) 103534
  %doi:10.1103/PhysRevD.86.103534
  [arXiv:1205.0868 [astro-ph.CO]].
  %%CITATION = doi:10.1103/PhysRevD.86.103534;%%
  %50 citations counted in INSPIRE as of 17 Apr 2019
  


  

  
\end{thebibliography}
\end{document}